\documentclass{isprs} 
\usepackage{subfigure}
\usepackage{setspace}
\usepackage{geometry} 
\usepackage{epstopdf}
\usepackage[labelsep=period]{caption}  
\usepackage[british]{babel} 
\usepackage[hang]{footmisc}
\usepackage{color}
\usepackage{hyperref}
\usepackage{mathtools}

\usepackage{booktabs}
\usepackage{amsmath}
\usepackage{amssymb}
\usepackage{tabularx}
\usepackage{multirow}
\usepackage{xcolor}

\usepackage{xurl}

\usepackage[authoryear]{natbib}

\begin{document}

\title{Country-wide, high-resolution monitoring of forest browning with Sentinel-2}
\date{}


\author{Samantha Biegel\textsuperscript{1}, David Brüggemann\textsuperscript{2}, Francesco Grossi\textsuperscript{3}, Michele Volpi\textsuperscript{2}, Konrad Schindler\textsuperscript{1}, Benjamin D. Stocker\textsuperscript{4, 5}}

\address{\textsuperscript{1 }Photogrammetry \& Remote Sensing, ETH Zurich and ETH AI Center - samantha.biegel@ai.ethz.ch, schindler@ethz.ch\\
\textsuperscript{2 }Swiss Data Science Center, ETH Zürich and EPFL - david.bruggemann@datascience.ch, michele.volpi@sdsc.ethz.ch\\
\textsuperscript{3 }Forest and Soil Ecology, Swiss Federal Institute for Forest, Snow and Landscape Research WSL,\\ 8903 Birmensdorf, Switzerland - francesco.grossi@wsl.ch \\  
\textsuperscript{4 }Institute of Geography, University of Bern, Hallerstrasse 12, 3012 Bern, Switzerland - benjamin.stocker@unibe.ch\\
\textsuperscript{5 }Oeschger Centre for Climate Change Research, University of Bern, Falkenplatz 16, 3012 Bern, Switzerland}



\abstract{Natural and anthropogenic disturbances are impacting the health of forests worldwide. Monitoring forest disturbances at scale is important to inform conservation efforts. Here, we present a scalable approach for country-wide mapping of forest greenness anomalies at the 10 m resolution of Sentinel-2. Using relevant ecological and topographical context and an established representation of the vegetation cycle, we learn a predictive quantile model of the normalised difference vegetation index (NDVI) derived from Sentinel-2 data. The resulting expected seasonal cycles are used to detect NDVI anomalies across Switzerland between April 2017 and August 2025. Goodness-of-fit evaluations show that the conditional model explains 65\% of the observed variations in the median seasonal cycle. The model consistently benefits from the local context information, particularly during the green-up period. The approach produces coherent spatial anomaly patterns and enables country-wide quantification of forest browning. Case studies with independent reference data from known events illustrate that the model reliably detects different types of disturbances.}


\maketitle


\section{Introduction}
\label{sec:introduction}

Forest ecosystems cover about 30\% of the global land surface and provide a variety of ecosystem services to society, such as carbon sequestration, water regulation, and the provision of timber \citep{Bonan2008}. The function and integrity of forest ecosystems is increasingly disturbed due to intensifying climate change, disturbances through climatic extreme events and pest outbreaks \citep{Hoegh-Guldberg2018, Liu2023a, Hermann2023}. Climatic disturbances such as droughts can cause reduced photosynthesis and growth and can lead to leaf discolouration, premature leaf shedding, and mortality \citep{Brun2020}. Monitoring disturbance impacts at scale is important to inform and drive forest conservation and adaptation efforts.

In optical satellite remote sensing observations, forest stress responses can be observed as abnormal browning \citep{Liu2023a, Fuentes2025, deJong2012}. The \emph{normalised difference vegetation index} (NDVI) links the measured reflectance spectra in the optical and near-infrared range to leaf pigments, vegetation activity and health \citep{Field1995, Gong2024, Beck2006}, and is commonly used to measure vegetation greenness as well as negative anomalies of it, i.e., browning. Time series of NDVI follow a natural seasonal pattern determined by the vegetation phenology \citep{Caparros-Santiago2021}. Therefore, the concept of \textit{normal} vegetation greenness and its \textit{anomalies} is inherently relative to the seasonal phases determined by prevailing climatic conditions throughout the year. Due to the close relationship between vegetation phenology and climate \citep{Richardson2013, Vitasse2011}, variations in the seasonal timing of phenological events can occur at small spatial scales if local topography causes steep climatic gradients, e.g., across elevation, slopes of varying aspects, or variations in root zone moisture availability and vegetation and groundwater connectivity \citep{Fan2019, Vitasse2021}. Additionally, spatial variations of the greenness cycle are affected by the distribution of species with varying phenological responses and strategies \citep{Vitasse2009, Caparros-Santiago2021}.

Reconstructing an annual vegetation phenological cycle establishes a baseline for anomaly detection. \citet{Chavez2022} have introduced a method to estimate the reference phenological curve via kernel density estimation. While they show the utility of these curves for anomaly detection, the method does not scale to large areas because of its computational cost and the independent treatment of each location. Other work has largely concentrated on targeted, event-specific mapping of vegetation browning caused by known disturbances \citep{Sturm2022, Brun2020, Fernandez-Carrillo2020}. At present, there is still a lack of large-scale mapping efforts that monitor abnormal browning at the high spatial and temporal resolutions needed to reflect the heterogeneity of climate impacts on forests. 

In this work, we present a scalable approach to mapping forest anomalies across Switzerland from April 2017 to August 2025, using Sentinel-2 imagery \citep{Drusch2012}. Instead of learning a separate NDVI baseline for every location, we model the expected NDVI seasonal cycle as a function of vegetation and terrain covariates (such as forest height, tree species, elevation, and slope). Conditional quantile models capture how NDVI typically varies for landscapes with similar ecological and topographical characteristics, providing a consistent and spatially transferable representation of seasonal greenness dynamics. 
Using a well-established functional representation of the phenological greenness cycle, we further constrain the model to ecologically plausible seasonal behaviour. Deviations from the expected cycle are then identified as anomalies, which we map across forest areas of Switzerland for each Sentinel-2 acquisition in our nine-year study. In addition to detecting and mapping anomalies, our approach provides estimates of phenological parameters at 10 m ground sampling distance (GSD). In a set of experiments, we assess the model's goodness of fit to NDVI and illustrate how the estimated anomalies capture impacts of known events, including a drought, a forest fire, a storm, a clear-cut harvest, and a beetle outbreak. The code and interactive content are available at \url{https://github.com/SamanthaBiegel/s2-forest-browning-monitoring}.

\section{Related work}
\label{sec:related-work}

\textbf{Land surface phenology.} Land surface phenology (LSP) studies seasonal patterns in plant states, often through vegetation indices derived from satellite imagery \citep{Caparros-Santiago2021}. It is closely related to vegetation anomaly detection, due to the link between shifts in phenophases and climate-induced disturbances. \citet{Forkel2015a} highlighted the uncertainties that come with the determination of phenology metrics, in particular when estimating at coarse resolutions. \citet{Koch2025} showed that phenology metrics extracted from Sentinel-2 images for beech and spruce only slightly differ from ground-based observations, and that they capture similar trends and inter-annual patterns. \citet{Klosterman2014} used a sigmoid representation to estimate seasonal changes and phenological transition dates from MODIS imagery. They found a good agreement with ground-based observations for the estimated leaf-out date, but weaker agreement later in spring, in particular in locations with heterogeneous land cover---a limitation that could potentially be resolved with satellite data of higher resolution. \citet{Richardson2018} also compared ground-based phenology observations with MODIS-derived dates from \citep{Zhang2003}, and found a similar high level of agreement. Again, the coarse resolution of MODIS imagery emerged as a major challenge in heterogeneous landscapes \citep{Richardson2018}. \citet{Bolton2020} and \citet{Tran2023} presented work on LSP prediction at higher spatial resolution using Harmonized Landsat and Sentinel-2 data. The approach by \citet{Tran2023} requires the integration of remote sensing data with ground-based observations, making it challenging to scale. Like other work, \citet{Bolton2020} pointed out challenges related to phenology modelling of evergreen vegetation even at 30 m resolution. In line with the multi-sensor approaches, \citet{Kosczor2022} fit double logistic curves to combined Landsat and Sentinel-2 data to estimate annual phenology metrics. The separate annual fit is sensitive to inter-annual variations of the climate, which makes it challenging to reliably detect deviations from the normal vegetation cycle. 

\textbf{Vegetation disturbance mapping.} Other efforts to map anomalies in vegetation largely focus on specific climatic disturbances. \citet{Brun2020} used change point analysis to determine areas of early wilting during the summer drought of 2018 in Switzerland. \citet{Sturm2022} compared the median \emph{normalised difference water index} (NDWI, as a proxy for vegetation water content) for the month of August between pre-drought, drought and post-drought years of the same drought event. They found that for 4.3\% of the Swiss forest area, NDWI declined by at least 10\%. \citet{Turner2023} utilised vegetation anomalies to quantify hurricane impact, relative to a reference period. At a larger spatial scale, annual forest disturbance maps have been generated for large study areas in Germany \citep{Reinosch2025}, Italy \citep{Francini2022} and Hungary \citep{Molnar2024} using composites of Sentinel-2 imagery. The aggregation of the signal to an annual time scale makes it challenging to distinguish between individual events and to identify drivers of forest disturbances. Similarly to our work, \citet{Low2020} employ phenology modelling to detect forest disturbances in Austria at the temporal resolution of Sentinel-2. They treat every pixel independently and estimate percentiles of the seasonal cycle per pixel based on two years of data, a relatively short period that makes it difficult to reliably detect anomalies. 

With the present work, our aim is to contribute to these efforts and introduce a methodology that is agnostic to the type and extent of a disturbance, and makes use of the full spatial and temporal resolution of Sentinel-2. Unlike other approaches that primarily use per-location historical data as a reference, we develop a predictive model based on environmental covariates to establish the normal state of vegetation, i.e., the mean seasonal cycle of NDVI.


\section{Methods}
\label{sec:methods}

\textbf{Data.} We have compiled a dataset that covers all forests in Switzerland, and all available Sentinel-2 images between 01-04-2017 and 31-08-2025. We derive NDVI at 10 m GSD from the red and near infrared reflectance bands as provided in the swissEO S2-SR dataset \citep{swisstopo_swissEO}. That dataset consists of Copernicus images with custom preprocessing, including co-registration optimised for Switzerland, cloud masking and terrain shadow and cloud shadow detection \citep{Pasquarella2023}. Snow was identified as recommended by the data provider, by thresholding the \emph{normalised difference snow index} derived from the same Sentinel-2 images at 0.43. For that purpose, the shortwave infrared band was bilinearly resampled to 10 m GSD. All NDVI observations were filtered using an outlier mask (NDVI between -0.1 and 1), the derived snow mask, and the cloud, cloud shadow, terrain shadow and data availability masks provided as part of the swissEO dataset. We also created a forest mask by identifying fully or partially forested pixels according to the 1 meter GSD Swiss Habitat Map \citep{Price2024}.

Alongside the NDVI time series per forest pixel, we created a collection of features considered relevant for the seasonality of the NDVI, see Table~\ref{tab:data}. Elevation, slope, northness and eastness\footnote{Derived from aspect as $n\!=\!-\cos(a\cdot\tfrac{\pi}{180})$ and $e\!=\!-\sin(a\cdot\tfrac{\pi}{180})$}, terrain ruggedness index and roughness were computed from the 2 m SwissAlti3D Digital Elevation Model \citep[DEM,][]{swisstopo_swissalti3d} and resampled to 10 m by averaging. Terrain wetness index (TWI), mean curvature, profile curvature and plan curvature were computed from the 10 m DEM. To obtain the TWI, we first filled depressions in the DEM using WhiteBoxTools \citep{whiteboxtools_2023}. The flow accumulation grid and specific catchment area were then estimated using the D8 flow model \citep{OCallaghan1984} as implemented in TauDEM \citep{taudem_5}. WhiteBoxTools was then used for the final step of TWI computation. With the exception of TWI, all DEM operations were performed with GDAL \citep{Rouault2025}.

Moreover, we included the median annual vegetation height for the years 2017 to 2023 \citep{Ginzler2021}, the forest mix rate \citep{Waser2025}, the estimated primary tree species \citep{Koch2024a}, and the habitat type \citep{Price2024}. The tree species map was reprojected to the Swiss national coordinate system with nearest neighbour resampling. The frequency distribution of habitat types for each Sentinel pixel was obtained by collecting the corresponding 100 pixels from the 1 m Swiss Habitat Map. For continuous features missing values were imputed (mean for vegetation height, median for forest mix rate), and all values were standardised to zero mean and unit standard deviation. For categorical features missing values were set to a separate category ``Unknown''.

\begin{table}[h]
\centering
\footnotesize
\begin{tabularx}{0.9\columnwidth}{ll}
\toprule
\textbf{Variable} & \textbf{Source} \\
\midrule
NDVI & Swisstopo / Copernicus \\
\midrule
Median Forest Height & \citet{Ginzler2021} \\
Forest Mix Rate & \citet{Waser2025} \\
Tree Species & \citet{Koch2024a}\\
Habitat Type & \citet{Price2024}\\
\midrule
Elevation & \multirow{10}{*}{SwissAlti3D DEM}\\
Slope &  \\
Eastness & \\
Northness & \\
Terrain Wetness Index & \\
Terrain Ruggedness Index$\qquad$ &\\
Mean Curvature & \\
Profile Curvature & \\
Plan Curvature & \\
Roughness & \\
\bottomrule
\end{tabularx}
\vspace{-0.5em}
\caption{Variables used in the NDVI seasonal cycle model.}
\label{tab:data}
\end{table}

\textbf{Model.} We represent the seasonal cycle of vegetation greenness as a double logistic function \citep{Beck2006}. Ecological studies have confirmed that tree phenology follows a corresponding regular temporal pattern \citep{Zhang2003}: From the appearance of leaves, there is a period of rapid growth (increase in greenness), followed by a period of relatively stable greenness. This behaviour fits a logistic model. The transition to dormancy follows a reverse pattern \citep{Zhang2003}. While the pattern is more pronounced in deciduous vegetation, the double logistic representation has been shown to be suitable also for NDVI curves in coniferous forests and in areas affected by snow \citep{Beck2006}. We define the NDVI curve as a function of phenological parameters in the following way:
{
\footnotesize
\begin{equation}
\begin{aligned}
f(t) = & \mathrm{NDVI}_{\min} + (\mathrm{NDVI}_{\max} - \mathrm{NDVI}_{\min}) \;\cdot \\ & \cdot \Big[
\sigma\!\left(
-2 \tfrac{2\,\mathrm{SOS} + g(\mathrm{MAT\!-\!SOS}) - 2t}{g(\mathrm{MAT\!-\!SOS})}
\right) \\
& \quad\; -\sigma\!\left(
-2 \tfrac{2\,\mathrm{SEN} + g(\mathrm{EOS\!-\!SEN}) - 2t}{g(\mathrm{EOS\!-\!SEN})}
\right) \Big]\;,
\end{aligned}
\label{eq:double_logistic}
\end{equation}
}%
where $t$ is the day of the year normalised by the year's length, $\sigma(\cdot)$ is the sigmoid function, $g(x) = \ln(1 + e^x)$ is the softplus function, $\mathrm{NDVI}_{\max}$ is the peak NDVI, $\mathrm{NDVI}_{\min}$ is the lowest NDVI, $\mathrm{SOS}$ is the start of the season (date of onset of photosynthetic activity), $\mathrm{MAT\!-\!SOS}$ is the duration of green-up (start of season until maturity), $\mathrm{SEN}$ is the start of senescence (date when photosynthetic activity starts to decline) and $\mathrm{EOS\!-\!SEN}$ is the duration from senescence until dormancy (date when photosynthetic activity is near zero). $\mathrm{SOS}$, $\mathrm{MAT}$, $\mathrm{SEN}$ and $\mathrm{EOS}$ are days of the year normalised by the year's length.

We learn conditional quantiles of NDVI to obtain its expected value range as a function of local vegetation characteristics and topography. A multilayer perceptron (MLP) is trained to predict the phenological parameters that define the quartile functions 
$f_{0.25}(t)$, $f_{0.5}(t)$ and $f_{0.75}(t)$ of the seasonal NDVI distribution. The double logistic model in Eq.~\eqref{eq:double_logistic} has six parameters, hence the network outputs 18 values\textemdash six for each of the three quantiles\textemdash to fit separate parameter sets for each quantile. The network includes embedding layers that map the two categorical inputs to continuous embedding vectors (length 4 for tree species, length 8 for habitat type), followed by a sequence of eight linear layers with ReLU activation (256 neurons per layer). A skip connection is added from the input to the fifth block to improve gradient flow \citep{He2016b}. Embeddings for each individual habitat category are learned from the sub-pixel frequencies. The habitat embedding vector for a pixel is the weighted mean of the learned habitat embeddings, with weights proportional to their frequencies within a Sentinel-2 pixel. The embedding vectors therefore encode the mix of habitats present at a pixel. They are concatenated with the continuous features as input to the linear layers.

\textbf{Model training.} We train the network by minimising the pinball loss \citep{Koenker1978}, applied directly between NDVI observations and the estimated quantile functions $f_{0.25}(t)$, $f_{0.5}(t)$ and $f_{0.75}(t)$ at the corresponding day of year and location. To ensure stable quantile estimation throughout the year, we stratify samples by day of year and reweigh each time step to compensate for differences in the number of valid observations, giving each day equal influence during training. The network is trained with the AdamW optimiser \citep{Loshchilov2019} with a weight decay of 0.0001 and an initial learning rate of 0.005, which exponentially declines by a factor of 0.01 during training. The model is nudged towards periodicity with an additional penalty for (squared) differences between NDVI predictions at the beginning and end of the year ($t=0$ and $t=1$). A further loss penalty discourages curves that cross quantiles. The complete loss function reads
{
\footnotesize
\begin{align}
&\mathcal{L}_{\text{total}}=\notag\\
&\underbrace{
 \sum_{q\in\{l,m,u\}}
 \frac{1}{T}
 \left\{
 \sum_{t:\,y_t\ge f_q(t)}
 \!\!\!\!\!
 q\,\lvert y_t-f_q(t)\rvert
 +
 \!\!\!\!\!
 \sum_{t:\,y_t<f_q(t)}
 \!\!\!\!(1-q)\,\lvert y_t-f_q(t)\rvert
 \right\}
 }_{\text{Quantile pinball losses}}
\notag\\[6pt]
&\quad
+\,\lambda_{\text{per}}\,
\underbrace{
\sum_{q\in\{l,m,u\}}
\big(
f_q(t_0)
-
f_q(t_1)
\big)^2
}_{\text{Periodicity constraint}}
\notag\\[6pt]
&\quad
+\,\lambda_{\text{nc}}\,
\underbrace{
\mathbb{E}_{t}\!\left\{
\sum_{(i,j)\in\mathcal{P}}
\max\!\big(0,\,f_i(t)-f_j(t)\big)
\right\}
}_{\text{Non-crossing constraint}}
\label{eq:total_loss}
\end{align}
}%
where $y_t$ is the observed NDVI value at time $t$, $f_q(t)$ is the model predicted $q$-quantile of NDVI at time $t$, $l$ is the lower quartile ($q=0.25$), $m$ is the median ($q=0.5$), $u$ is the upper quartile ($q=0.75$), $\mathcal{P}=\{(0.25,0.5), \allowbreak (0.5,0.75), \allowbreak (0.25,0.75)\}$ and $\lambda_{\text{per}}=1$ and $\lambda_{\text{nc}}=10$ are scaling factors that balance the contributions of each component to the total loss.

We train and evaluate the model on the complete dataset of Swiss forest areas 2017--2025. Training was run for 20 epochs on the full dataset in batches of 1024 forest pixels. For fast data retrieval, those batches were drawn randomly from pre-shuffled chunks of 8192 pixels, with the chunks reshuffled at every epoch.

\begin{figure}[ht!]
\begin{center}
        \vspace{1em}
		\includegraphics[width=1.0\columnwidth]{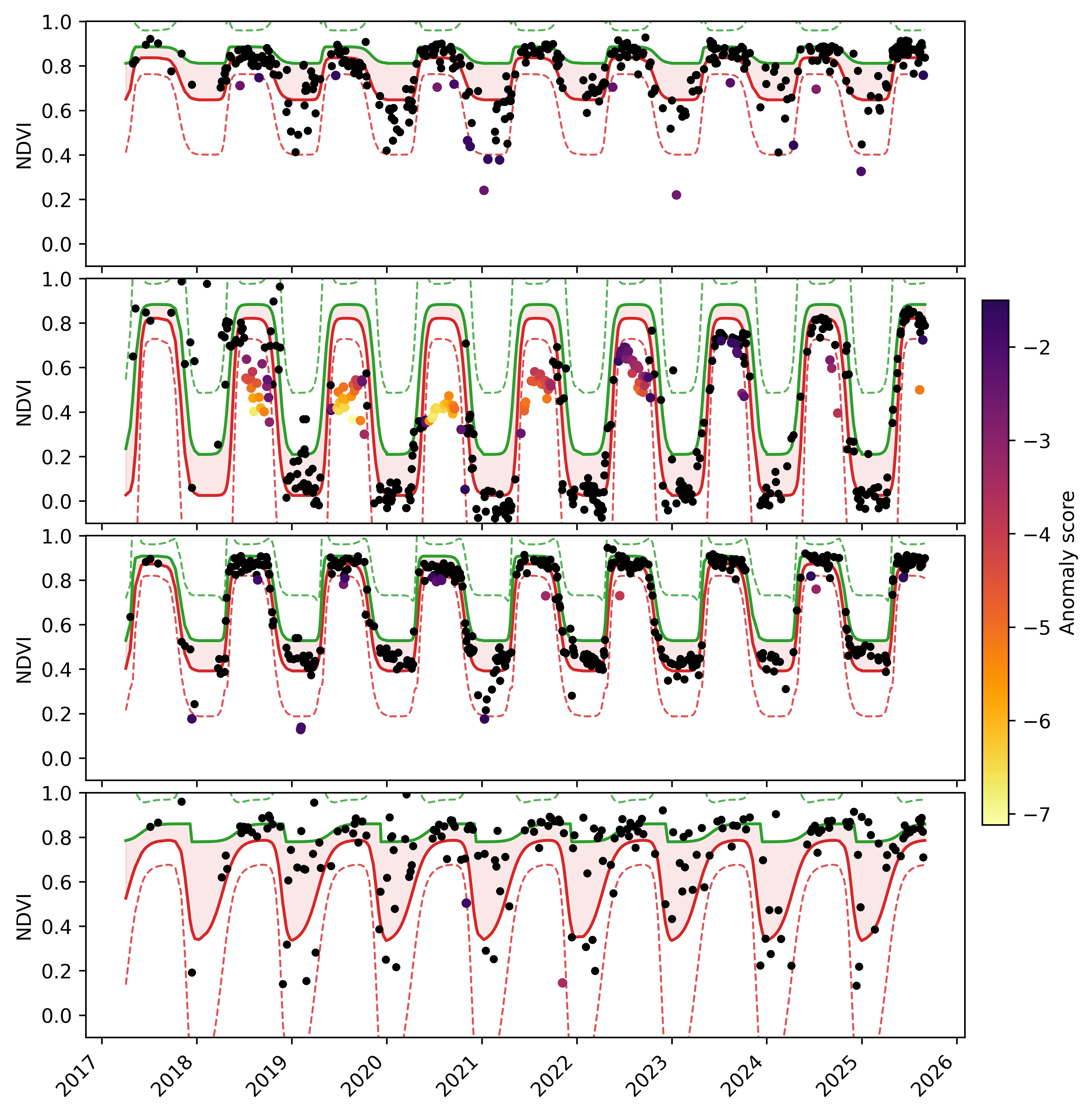}
	\caption{Exemplary predicted NDVI ranges and anomalies for 4 forest pixels. Green curves denote the upper quartiles ($f_{0.75}(t)$), red curves denote the lower quartiles ($f_{0.25}(t)$). Interquartile ranges are shaded in red. Dashed lines indicate the anomaly thresholds derived from those curves. NDVI observations that fall within the anomaly thresholds are coloured black, observations labelled as anomalies are coloured by their anomaly score, with lighter colour denoting more negative scores (larger deviations from the normal range).}
\label{fig:curves}
\end{center}
\end{figure}

\begin{figure}[ht!]
\begin{center}
		\includegraphics[width=1.0\columnwidth]{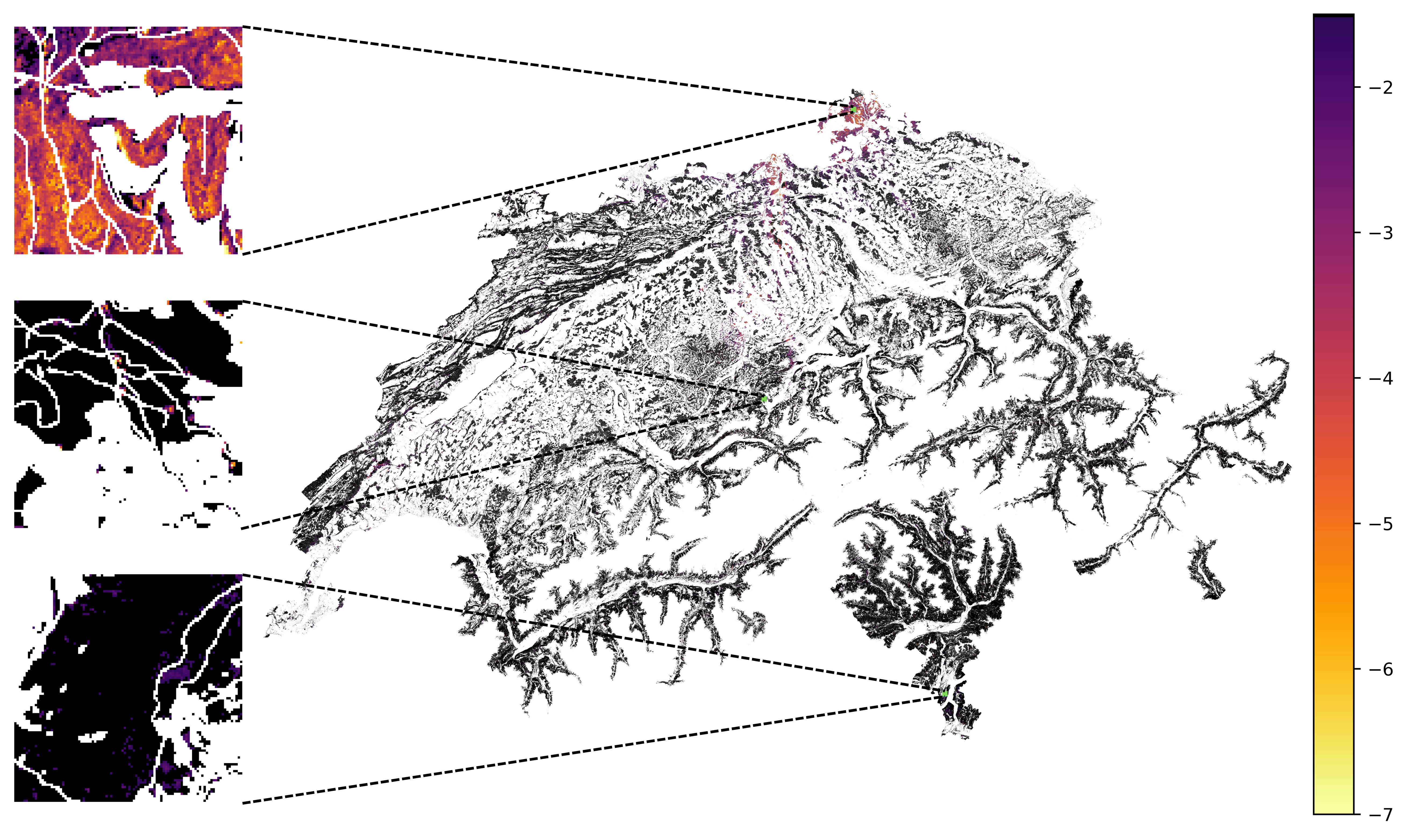}
	\caption{Example anomaly map for forest pixels based on NDVI observations from 22/08/2023 and 24/08/2023 combined. A pixel is flagged if it is detected as anomalous on either of the days. The three panels on the left show a magnified view of the anomalies at the indicated regions on the map. Normal pixels are coloured in black and anomalous pixels are shaded in red according to their anomaly score.}
\label{fig:anomaly_map}
\end{center}
\end{figure}

\textbf{Anomaly detection.} Having determined the interquartile range $\text{IQR}(t) = f_{0.75}(t) - f_{0.25}(t)$ for each pixel and each day $t$ of the year, we compute an anomaly score $(y_t - f_{0.25}(t)) / \text{IQR}(t)$ and define (negative) anomalies as NDVI observations with a score below -1.5, i.e., their value lies below the negative anomaly threshold $f_{0.25}(t) - 1.5 \; \text{IQR}(t)$. Figure~\ref{fig:curves} shows examples of estimated NDVI curves and anomalies for selected pixels. For every Sentinel overpass, we generate maps of detected anomalies. An example is shown in Figure~\ref{fig:anomaly_map}. Our focus is on negative anomalies, as these represent forest disturbances with browning impact. We note that the proposed method is less sensitive to positive anomalies due to NDVI values being generally high in Switzerland during summer and having a large spread during winter, which raises the upper anomaly threshold close to (or even beyond) the theoretical maximum NDVI of 1.

\section{Experiments}
\label{sec:results}

\subsection{Goodness of fit}

We compare the model against two baselines:
(1) A country-wide day of year-independent quantile, estimated by pooling NDVI observations across all spatial locations and time steps (\emph{global baseline});
(2) A country-wide per-day quantile, estimated by pooling NDVI observations across all spatial locations for each day of the year (\emph{climatology baseline}).
The global baseline corresponds to a model that uses the overall empirical quantile as constant prediction, without using any spatial or temporal information. The climatology baseline captures the overall historical seasonal cycle without using any spatial information. These two baselines help distinguish whether benefits to predictive accuracy occur due to spatial or temporal information.

{
\setlength{\tabcolsep}{3pt}
\begin{table*}
\centering
\footnotesize
\begin{tabular}{lcccccccccccccccc}
\toprule
\multirow{2}{*}{\textbf{Model}} & \multicolumn{3}{c}{Pinball loss} && \multicolumn{3}{c}{D$^2$} && \multicolumn{3}{c}{Quantile coverage} && \multirow{2}{*}{MAE} & \multirow{2}{*}{RMSE} & \multirow{2}{*}{Bias} & \multirow{2}{*}{R$^2$} \\
\cmidrule{2-4}
\cmidrule{6-8}
\cmidrule{10-12}
& q = 0.25 & q = 0.5 & q = 0.75 && q = 0.25 & q = 0.5 & q = 0.75 && q = 0.25 & q = 0.5 & q = 0.75 &&&&& \\
\midrule
Global baseline & 0.081 & 0.080 & 0.048 && 0.0 & 0.0 & 0.0 && 0.248 & 0.500 & 0.752 && 0.16 & 0.23 & 0.08 & -0.15 \\
Climatology baseline & 0.048 & 0.054 & 0.039 && 0.41 & 0.32 & 0.19 && 0.250 & 0.500 & 0.750 && 0.11 & 0.16 & 0.02 & 0.46 \\
Conditional model & 0.037 & 0.042 & 0.031 && 0.55 & 0.47 & 0.36 && 0.245 & 0.497 & 0.748 && 0.08 & 0.13 & 0.01 & 0.65 \\
\bottomrule
\end{tabular}
\caption{Quantile model goodness of fit. The \emph{global baseline} uses a single, country-wide empirical quantile. The \emph{climatology baseline} uses a country-wide quantile per day of the year. MAE, RMSE, bias and R$^2$ are for per-day median values. ``Bias'' is the difference between the mean of model predictions and the mean of observations.}
\label{tab:metrics}
\end{table*}
}

Table~\ref{tab:metrics} summarises the goodness of fit across several metrics: D$^2$ (fraction of pinball loss explained), quantile coverage, mean absolute error (MAE), root mean squared error (RMSE), bias, and R$^2$. The conditional model achieves its highest D$^2$ score for the 0.25 quantile (0.55), while the score is lowest for the 0.75 quantile. This behaviour reflects the comparatively low loss of the global baseline at the 0.75 quantile: the spatially pooled empirical upper quantile already approximates the data well, leaving less room for improvement by the conditional model. The comparison of D$^2$ between the model and baselines reveals that conditioning on local context leads to a significantly better goodness of fit compared to a single seasonal baseline. E.g., the conditional model explains 47\% of the pinball loss for the median prediction, while the climatology baseline explains 32\% of the pinball loss. Coverage values indicate that the model attains the target quantile levels across the distribution.

\begin{figure}[ht!]
\begin{center}
	\includegraphics[width=1.0\columnwidth]{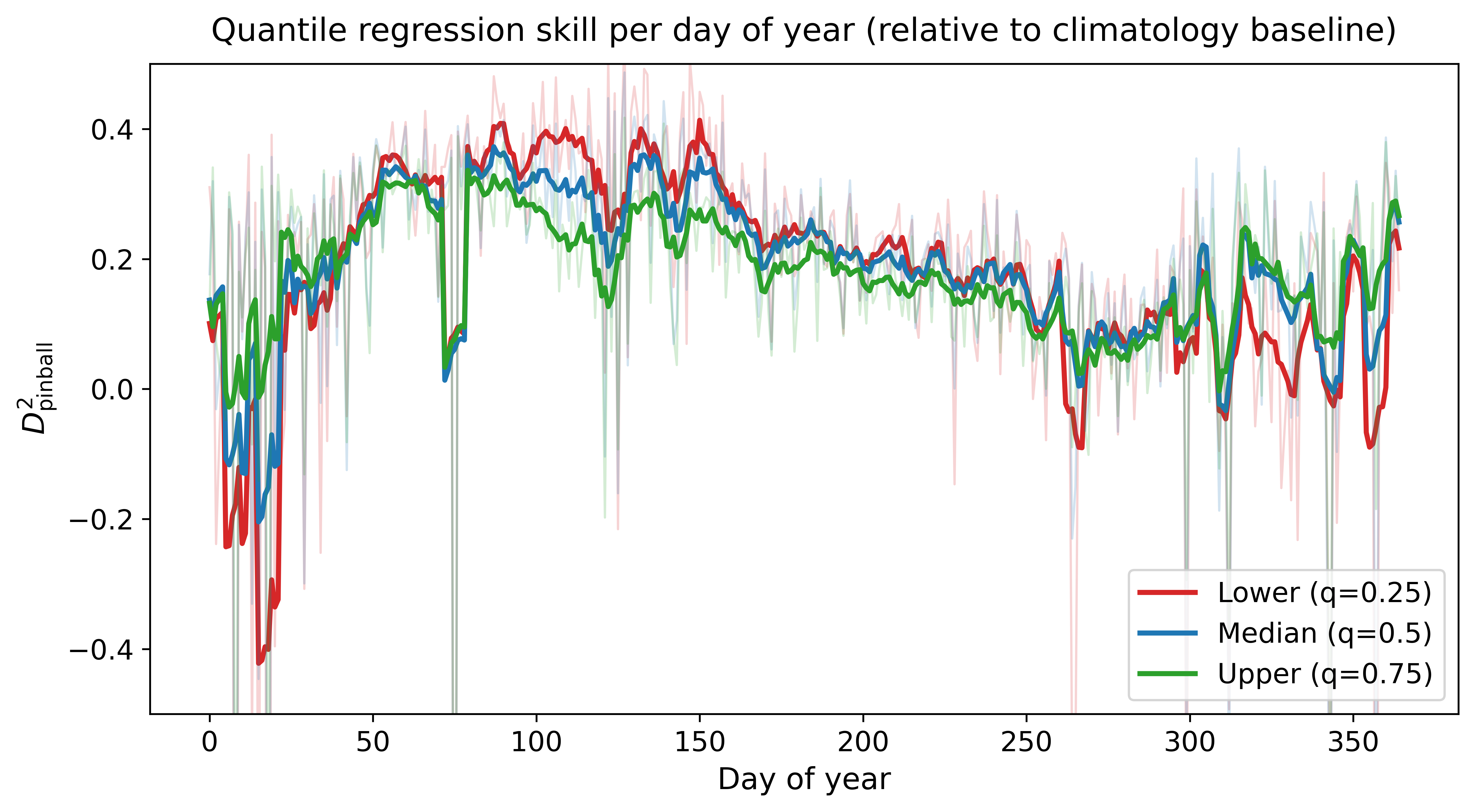}
	\caption{D$^2_{\text{pinball}}$ per day of the year for each of the predicted quartiles. The daily values are shown alongside the 7-day rolling mean. For each day of the year, the value of D$^2_{\text{pinball}}$ expresses the fraction of pinball loss explained when compared to a model that uses the empirical quantile at that day of the year (the \emph{climatology baseline}). A value of 0 corresponds to a model that predicts the same value at all locations per day of the year.}
\label{fig:score_doy}
\end{center}
\end{figure}

Figure~\ref{fig:score_doy} shows the D$^2_{\text{pinball}}$ score across all pixels per day of the year. The predictive skill of the model (relative to the climatology baseline) is highest during early spring, between day 50 and day 150 of the year. During winter, the model has limited benefit, and on some days even worse performance than the unconditional quantile (climatology baseline). In winter, baseline NDVI is low and exhibits little variation among vegetation types and topographical features, consequently the additional features are less informative. In contrast, during summer, the photosynthetic activity strongly varies depending on local growing conditions, leading to variations in greenness that can be predicted better when including local context.

\subsection{Anomalies}

\begin{figure}[ht!]
\begin{center}
	\includegraphics[width=1.0\columnwidth]{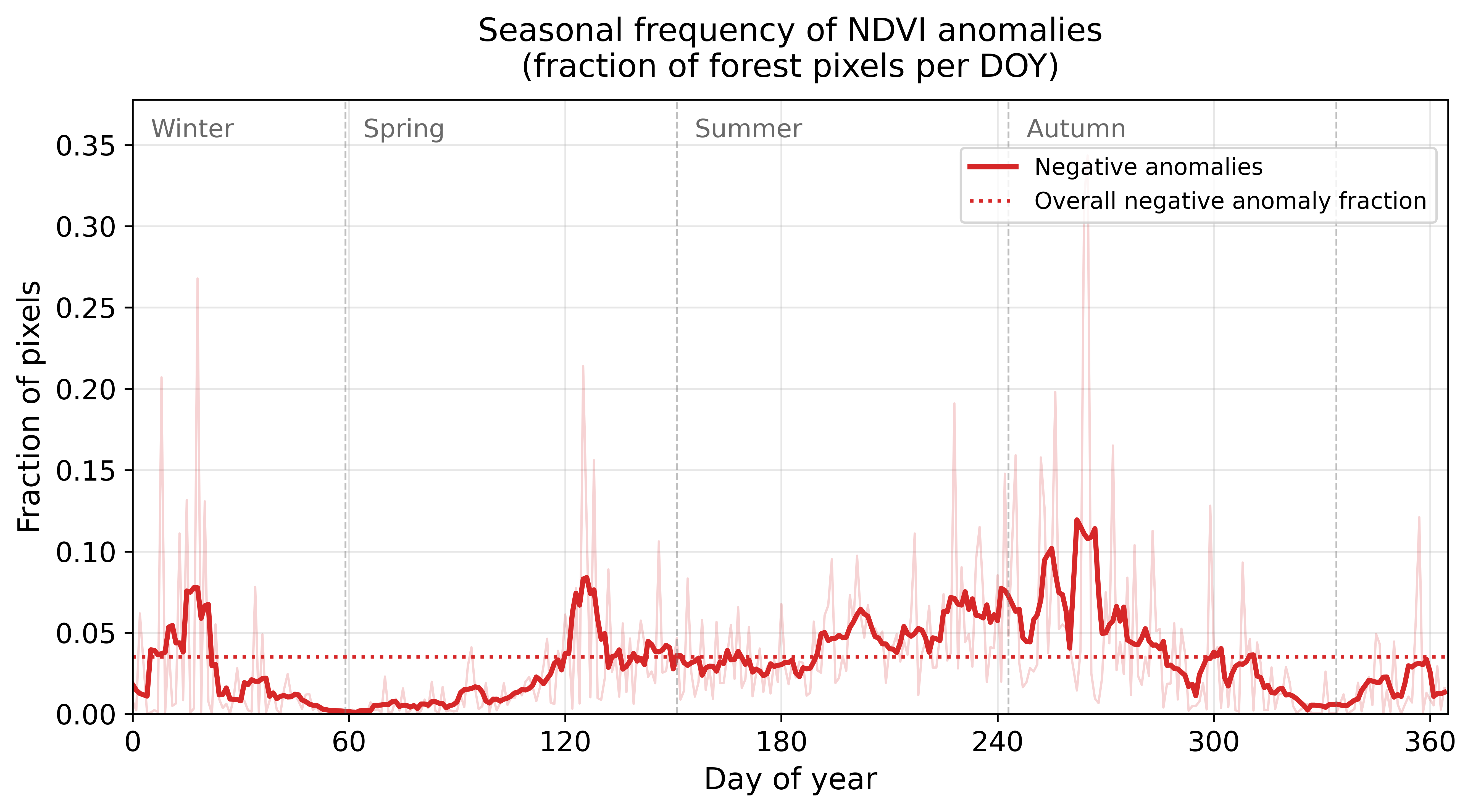}
	\caption{Daily fraction of forest pixels affected by negative NDVI anomalies. The negative anomaly fractions are computed per day of the year and smoothed with a 7-day rolling average. The dashed horizontal line indicates the overall fraction of anomalies detected at forest pixels. Fractions are relative to the number of valid observations at a pixel. The vertical lines indicate the season starts on March 1 (spring), June 1 (summer), September 1 (autumn) and December 1 (winter).}
\label{fig:anomalies_doy}
\end{center}
\end{figure}

We detected anomalous browning at 3.5\% of all pixel observations. The browning varies throughout the year, as shown in Figure~\ref{fig:anomalies_doy}. High proportions of anomalous pixels during the peak growing season correspond to browning events, whereas anomalies at the beginning or end of the growing seasons can be interpreted in several ways: they could hint at events that cause later onset of the green-up or early wilting, or they could be detections of inter-annual variations in phenological dates. Overall, the fraction of negative anomalies is highest between the end of spring and the middle of autumn, with peaks during late spring and late summer-early autumn. The peaks could indicate high inter-annual variability and could also be a reflection of the choice to constrain the vegetation cycle to a double logistic shape. This leads to slightly more anomalies during late summer due to the slight decline of vegetation greenness over summer. Finally, the peak during winter is likely caused by undetected snow cover.

\begin{figure*}[t]
\begin{center}
		\includegraphics[width=1.0\textwidth]{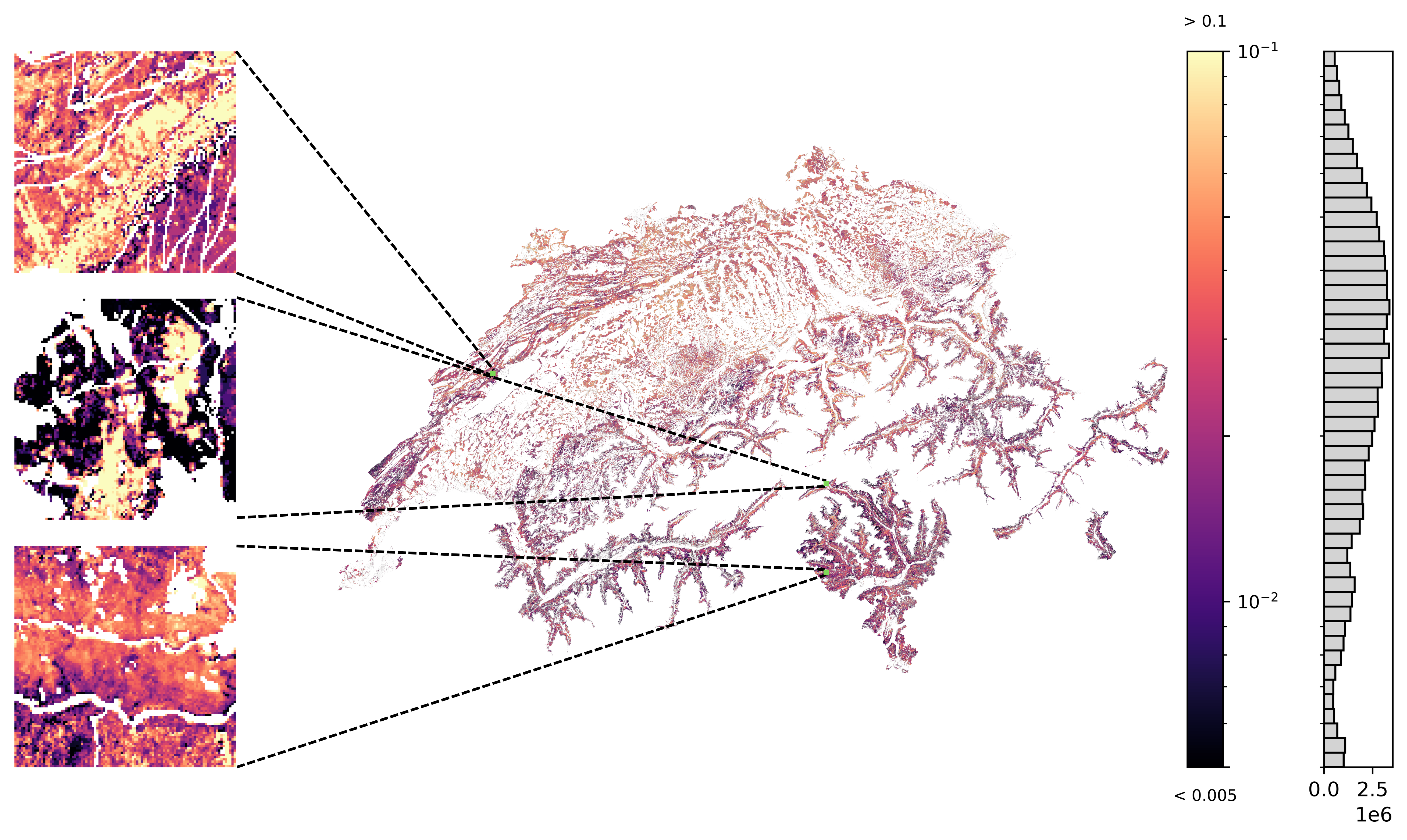}
	\caption{Fraction of observations per pixel affected by negative NDVI anomalies (``browning'') over the entire observation period. Fractions are relative to the number of valid observations at a pixel. The histogram on the right shows the distribution of anomaly fraction values across all pixels. The three panels on the left show a magnified view of the anomaly fractions at the indicated regions on the map.}
\label{fig:anomaly_frac_map}
\end{center}
\end{figure*}

The level of abnormal browning also varies spatially. Figure~\ref{fig:anomaly_frac_map} shows the fraction of negative anomalies (relative to the number of valid observations at a given forest pixel). Most pixels have around 3 to 5\% of anomalies. Only $\approx$5\% of pixels have no anomalies at all during the observation period. 1\% of pixels have over 22\% of anomalous observations. It appears that high-altitude areas have a relatively high amount of anomalies. They are often covered by snow during the winter, such that many NDVI observations have to be filtered out. Since more anomalies are found during the summer (Figure~\ref{fig:anomalies_doy}), the per-pixel fractions appear elevated compared to areas with regular, year-round observations.

\subsection{Case studies}

We relate the detected anomalies to five known events in Switzerland that impacted forests within our study period. 
For all cases, we select a confirmed affected area and a control area and evaluate anomaly scores and the fraction of anomalous pixels before, during and after the event for both areas separately. For each Sentinel-2 acquisition, we take the median of the anomaly scores as well as the 25th and 75th percentile among all pixels within each area. For each observed date, we also take the fraction of anomalous pixels according to the model, using the standard anomaly threshold of -1.5.

The first example includes an area (652 pixels) known to be affected by a forest fire near Bitsch on 17-07-2023. The second example concerns a drought and covers an area (1100 pixels) affected by early wilting due to dry conditions in the summer of 2018 \citep{Brun2020}. The third example is an area (7518 pixels) affected by storm Burglind on 03-01-2018, and for which reference areas have been delineated in the field \citep{Jiang2023a}. The clearcut case corresponds to a routine timber harvest (250 pixels) near the town of Airolo in 2021. Finally, the beetle case (498 pixels) corresponds to an outbreak near Lake Zürich. For the forest fire, clearcut and beetle outbreak, we visually selected affected areas using the SWISSIMAGE orthophoto mosaic \citep{swisstopo_swissimage}. As reference areas, we use a randomly selected set of pixels outside the delineated event for the drought case, and visually unaffected control areas based on the SWISSIMAGE mosaic for the other events.

\begin{figure*}[ht!]
\begin{center}
		\includegraphics[width=1.0\textwidth]{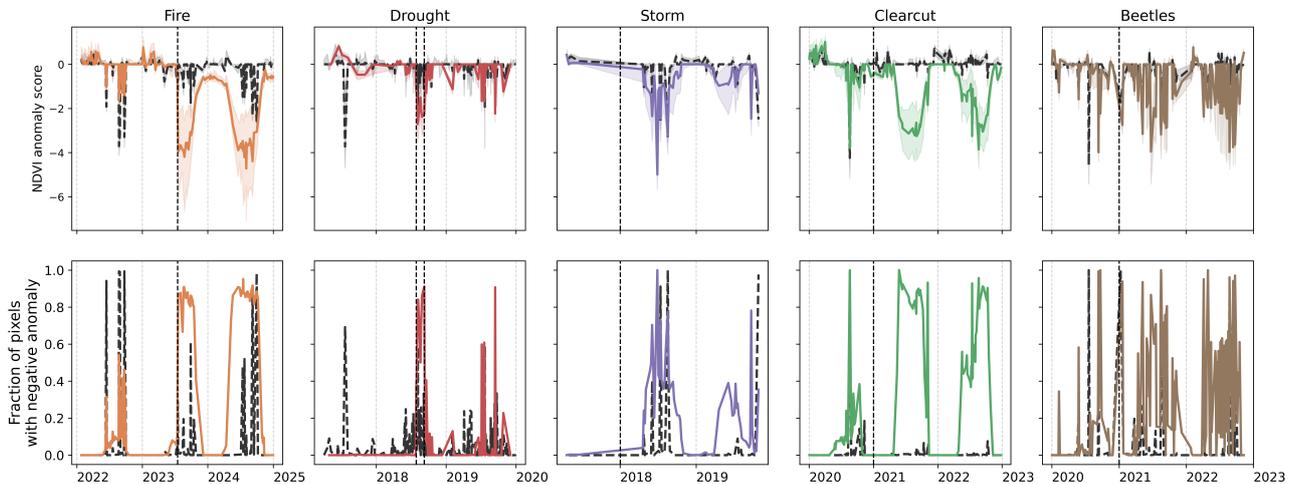}
        \caption{Aggregated anomaly scores (top row) and fraction of anomalous pixels (bottom row) for each of the five cases (one case per column). Coloured curves denote the cases and dashed grey lines denote the reference areas. The vertical dashed lines indicate the date of the event for the fire and storm, the start and end of drought impact determined with the anomaly data for the drought event, and the start of the year for the clearcut harvest and beetle outbreak. In addition to the median, the 25th and 75th percentiles of the anomaly scores are shown by the shaded area.}
\label{fig:cases}
\end{center}
\end{figure*}

For each of the five case studies, Figure~\ref{fig:cases} shows the average anomaly score (top row) and the fraction of anomalous pixels (bottom row) per acquisition day. As expected, the highest response is in all cases observed during the growing season. For the fire, over 75\% of the pixels were identified as anomalous during the period immediately after the event and also in the two following years. The median anomaly score fell below -4 during the post-fire period. The clearcut case shows a similarly strong anomaly response with at least 75\% of pixels detected as anomalies and the median anomaly score going down to -3 during the growing season. For the drought, the onset of forest browning was detected on 29-07-2018, when the fraction of anomalous pixels jumped to 37\%, compared to 0\% just three days earlier. Over 50\% of the pixels stayed anomalous until 09-09-2018. The storm and beetle cases show anomaly scores going down to -4 and also peak near 100\% anomalous pixels. Even the control areas show some peaks in these cases, but they appear relatively temporary compared to the consistent detections over time for the affected areas. In the fire, clearcut and beetle cases, significant browning impacts can still be observed during the post-event year, indicating that the disturbance may have caused tree mortality in these areas. 

\subsection{Model behaviour}

\begin{figure*}[ht!]
\centering
		\includegraphics[width=0.71\textwidth]{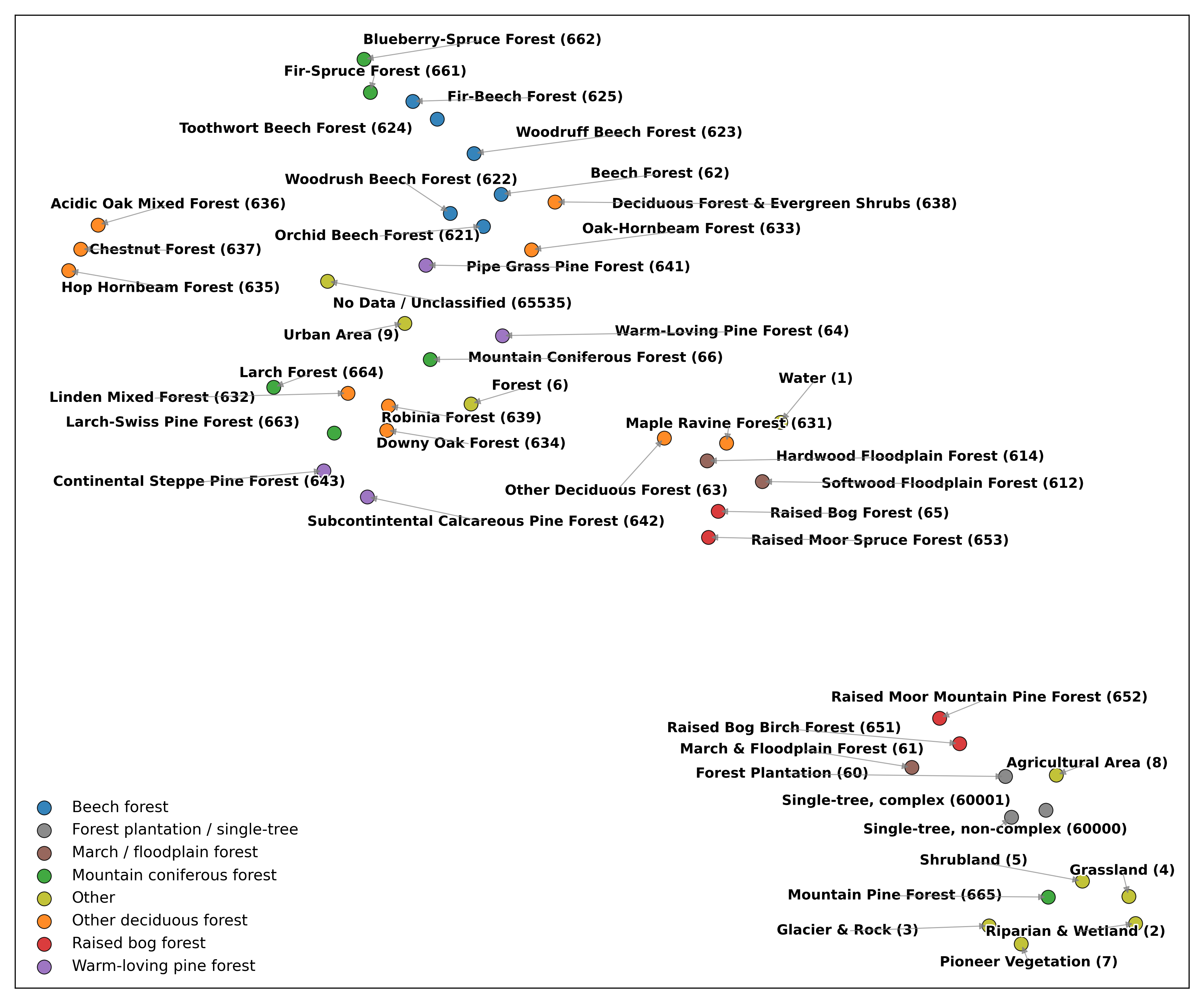}
	\caption{Visualisation of forest embeddings, projected to two dimensions with $t$-SNE. Points are labelled with their TypoCH habitat names and associated hierarchical codes, and coloured according to the highest hierarchy level.}
\label{fig:embeddings}
\end{figure*}

The Habitat Map uses a hierarchical encoding with up to three levels, following the TypoCH habitat classification \citep{Delarze2008}. Our model learned embeddings for the habitat encodings. To understand to what extent the embeddings reflect the properties of the corresponding habitat types, we visualise them in 2D using $t$-distributed Stochastic Neighbour Embedding \citep[$t$-SNE,][]{Maaten2008}. The resulting plot reveals plausible structures (Figure~\ref{fig:embeddings}). More drought-adapted forests are on the left side of the diagram, while forests that prefer moist climates are on the right, floodplain and bog forests form a clear cluster. Non-forest ecosystems appear on the bottom right. Mountain coniferous forests are separated into a deciduous cluster (with larch trees) and an evergreen cluster at the top. While beech forests are deciduous, their dried leaves tend to remain on the branches during winter, which places them near the evergreen cluster. Below the beech forests, the region in the centre primarily contains deciduous forests.

\section{Discussion}
\label{sec:discussion}

The modelling approach presented here uses a relatively rigid shape of the yearly phenological cycle. The results indicate that the double logistic function can accurately describe the seasonal cycle of the temperate forests present in Switzerland due to the winter-dormant phenology with a single seasonal peak. Other regions and ecosystem types exhibit different phenological patterns that are influenced by different growing season constraints (e.g., seasonal water limitation), the presence of other plant functional types (e.g., annual herbaceous plants) or variations in land use (e.g., croplands). The modelling of greenness phenology with a constrained annual pattern might be too restrictive to generalise to such other systems. While our modelling approach should generalise well to similar systems (such as temperate forests in Europe), this is probably not the case for other biomes. 

An advantage of our approach is the use of a single model to obtain the parameters of the double logistic function and describe the vegetation cycle across large areas. While the double logistic model has been widely used in phenology modelling, these models are typically fitted separately per pixel and year. Our learned model makes use of patterns in features to predict per-pixel ranges of the greenness cycle. With separate models it is more challenging to robustly estimate NDVI ranges due to the limited time series length. On the other hand, the learned model requires the availability of ancillary data (i.e., features as described in this study) to enable prediction in other regions, which limits the scalability of the approach. 






\section{Conclusion}
\label{sec:conclusion}

We have introduced an approach to map unusual forest browning based on Sentinel-2 imagery. Our framework is scalable and makes it possible to monitor browning at high resolution (10 m) over large areas.
Its basis is a quantile model of the vegetation seasonal cycle, conditioned on local vegetation and topography features. In our application, covering 8.5 years of satellite data for Switzerland, the model explained 65\% of the observed spatial variations in the median seasonal cycle, across a variety of forest types and growth conditions including large topographic and moisture gradients.
We found that anomalies derived from the quantile model of the normal cycle exhibit plausible, coherent patterns and enable a consistent quantification of anomalous browning.
In five case studies with known, independently mapped disturbance events, we have verified that the proposed anomaly detection scheme is able to reliably identify forest disturbances of varying nature.

A remaining challenge is the large portion of missing NDVI observations due to unfavourable meteorological conditions. Frequent clouds, and in Switzerland also snow, cause long periods of missing data when detected, and noisy observations when not detected.
Our constrained, low-dimensional representation helps to nevertheless reconstruct the seasonal cycle and its expected deviations rather reliably, but can of course not avoid that anomalies are missed or incompletely detected due to data gaps. 
Future research into spatio-temporal understanding and imputation of greenness anomalies could give further insight into forest disturbance patterns, especially with respect to events that affect extended geographic regions.




\section*{Acknowledgements}

This research was supported by the ETH AI Center through a doctoral fellowship for Samantha Biegel and by the Swiss Data Science Collaborative Grant C24-03UBE -- DIMPEO.

{
	\begin{spacing}{1.17}
		\normalsize
		\bibliography{anomaly_ms}

@misc{swisstopo_swissEO,
  author       = {{swisstopo}},
  title        = {{swissEO S2-SR: Optical satellite data (Sentinel-2) for Switzerland}},
  year = 2024,
  howpublished          = {Federal Office of Topography, \url{https://www.swisstopo.admin.ch/en/satelliteimage-swisseo-s2-sr}},
  institution  = {Federal Office of Topography swisstopo, Switzerland}
}

@misc{swisstopo_swissimage,
  author       = {{swisstopo}},
  title        = {{SWISSIMAGE}},
  year = 2024,
  howpublished          = {Federal Office of Topography, \url{https://www.swisstopo.admin.ch/en/orthoimage-swissimage-10}},
  institution  = {Federal Office of Topography swisstopo, Switzerland}
}

@misc{whiteboxtools_2023,
  title        = {{WhiteboxTools}},
  author       = {Lindsay, John B.},
  year         = {2023},
  howpublished = {Geomorphometry and Hydrogeomatics Research Group, University of Guelph, \url{https://www.whiteboxgeo.com}}
}

@misc{taudem_5,
  title        = {{TauDEM 5: Terrain Analysis Using Digital Elevation Models}},
  author       = {Tarboton, David G.},
  organization = {Utah State University},
  year         = {2015},
  howpublished = {Utah State University, \url{https://hydrology.usu.edu/taudem/taudem5}},
}

@misc{swisstopo_swissalti3d,
  author       = {{swisstopo}},
  title        = {{swissALTI3D: High-Precision Digital Elevation Model of Switzerland}},
  year         = {2024},
  howpublished = {Federal Office of Topography, \url{https://www.swisstopo.admin.ch/en/height-model-swissalti3d}}
}

@article{Beck2006,
  title = {Improved monitoring of vegetation dynamics at very high latitudes: A new method using {MODIS} {NDVI}},
  shorttitle = {Improved monitoring of vegetation dynamics at very high latitudes},
  author = {Beck, Pieter S. A. and Atzberger, Clement and H{\o}gda, Kjell Arild and Johansen, Bernt and Skidmore, Andrew K.},
  year = 2006,
  journal = {Remote Sensing of Environment},
  volume = {100},
  number = {3},
  pages = {321--334}
}

@article{Bolton2020,
  title = {Continental-scale land surface phenology from harmonized Landsat 8 and Sentinel-2 imagery},
  author = {Bolton, Douglas K. and Gray, Josh M. and Melaas, Eli K. and Moon, Minkyu and Eklundh, Lars and Friedl, Mark A.},
  year = 2020,
  month = apr,
  journal = {Remote Sensing of Environment},
  volume = {240},
  pages = {111685},
  issn = {0034-4257},
  doi = {10.1016/j.rse.2020.111685},
  urldate = {2025-11-14}
}

@article{Bonan2008,
  title = {Forests and Climate Change: Forcings, Feedbacks, and the Climate Benefits of Forests},
  shorttitle = {Forests and Climate Change},
  author = {Bonan, Gordon B.},
  year = 2008,
  journal = {Science},
  volume = {320},
  number = {5882},
  pages = {1444--1449}
}

@article{Brun2020,
  title = {Large-scale early-wilting response of {Central} {European} forests to the 2018 extreme drought},
  author = {Brun, Philipp and Psomas, Achilleas and Ginzler, Christian and Thuiller, Wilfried and Zappa, Massimiliano and Zimmermann, Niklaus E.},
  year = 2020,
  journal = {Global Change Biology},
  volume = {26},
  number = {12},
  pages = {7021--7035}
}

@article{Caparros-Santiago2021,
  title = {Land surface phenology as indicator of global terrestrial ecosystem dynamics: A systematic review},
  shorttitle = {Land surface phenology as indicator of global terrestrial ecosystem dynamics},
  author = {{Caparros-Santiago}, Jose A. and {Rodriguez-Galiano}, Victor and Dash, Jadunandan},
  year = 2021,
  journal = {ISPRS Journal of Photogrammetry and Remote Sensing},
  volume = {171},
  pages = {330--347}
}

@article{Chavez2022,
  title = {npphen: An {R}-Package for Detecting and Mapping Extreme Vegetation Anomalies Based on Remotely Sensed Phenological Variability},
  shorttitle = {npphen},
  author = {Ch{\'a}vez, Roberto O. and Estay, Sergio A. and Lastra, Jos{\'e} A. and Riquelme, Carlos G. and Olea, Mat{\'i}as and Aguayo, Javiera and Decuyper, Mathieu},
  year = 2022,
  journal = {Remote Sensing},
  volume = {15},
  number = {1},
  pages = {73}
}

@article{deJong2012,
  title = {Trend changes in global greening and browning: contribution of short-term trends to longer-term change},
  shorttitle = {Trend changes in global greening and browning},
  author = {{de Jong}, Rogier and Verbesselt, Jan and Schaepman, Michael E. and {de Bruin}, Sytze},
  year = 2012,
  journal = {Global Change Biology},
  volume = {18},
  number = {2},
  pages = {642--655}
}

@book{Delarze2008,
  title = {{Lebensr{\"a}ume der Schweiz: {\"O}kologie -- Gef{\"a}hrdung -- Kennarten}},
  shorttitle = {Lebensr{\"a}ume der Schweiz},
  author = {Delarze, Raymond and Gonseth, Yves and Galland, Pierre and Delarze, Raymond},
  year = 2008,
  edition = {2nd},
  publisher = {hep Verlag}
}

@article{Drusch2012,
  title = {Sentinel-2: ESA's Optical High-Resolution Mission for GMES Operational Services},
  shorttitle = {Sentinel-2},
  author = {Drusch, M. and Del Bello, U. and Carlier, S. and Colin, O. and Fernandez, V. and Gascon, F. and Hoersch, B. and Isola, C. and Laberinti, P. and Martimort, P. and Meygret, A. and Spoto, F. and Sy, O. and Marchese, F. and Bargellini, P.},
  year = 2012,
  month = may,
  journal = {Remote Sensing of Environment},
  series = {The Sentinel Missions - New Opportunities for Science},
  volume = {120},
  pages = {25--36},
  issn = {0034-4257},
  doi = {10.1016/j.rse.2011.11.026},
  urldate = {2025-11-17}
}

@article{Fan2019,
  title = {Hillslope Hydrology in Global Change Research and Earth System Modeling},
  author = {Fan, Y. and Clark, M. and Lawrence, D. M. and Swenson, S. and Band, L. E. and Brantley, S. L. and Brooks, P. D. and Dietrich, W. E. and Flores, A. and Grant, G. and Kirchner, J. W. and Mackay, D. S. and McDonnell, J. J. and Milly, P. C. D. and Sullivan, P. L. and Tague, C. and Ajami, H. and Chaney, N. and Hartmann, A. and others},
  year = 2019,
  journal = {Water Resources Research},
  volume = {55},
  number = {2},
  pages = {1737--1772}
}

@article{Fernandez-Carrillo2020,
  title = {Monitoring Bark Beetle Forest Damage in {Central} {Europe}. A Remote Sensing Approach Validated with Field Data},
  author = {{Fernandez-Carrillo}, Angel and Pato{\v c}ka, Zden{\v e}k and Dobrovoln{\'y}, Lum{\'i}r and {Franco-Nieto}, Antonio and {Revilla-Romero}, Beatriz},
  year = 2020,
  month = nov,
  journal = {Remote Sensing},
  volume = {12},
  number = {21},
  pages = {3634}
}

@article{Field1995,
  title = {Global net primary production: Combining ecology and remote sensing},
  shorttitle = {Global net primary production},
  author = {Field, Christopher B. and Randerson, James T. and Malmstr{\"o}m, Carolyn M.},
  year = 1995,
  month = jan,
  journal = {Remote Sensing of Environment},
  series = {Remote Sensing of Land Surface for Studies of Global Chage},
  volume = {51},
  number = {1},
  pages = {74--88},
  issn = {0034-4257},
  doi = {10.1016/0034-4257(94)00066-V},
  urldate = {2026-03-25}
}

@article{Forkel2015a,
  title = {Codominant water control on global interannual variability and trends in land surface phenology and greenness},
  author = {Forkel, Matthias and Migliavacca, Mirco and Thonicke, Kirsten and Reichstein, Markus and Schaphoff, Sibyll and Weber, Ulrich and Carvalhais, Nuno},
  year = 2015,
  journal = {Global Change Biology},
  volume = {21},
  number = {9},
  pages = {3414--3435},
  issn = {1365-2486},
  doi = {10.1111/gcb.12950},
  urldate = {2026-03-19},
  langid = {english}
}

@article{Francini2022,
  title = {An open science and open data approach for the statistically robust estimation of forest disturbance areas},
  author = {Francini, Saverio and McRoberts, Ronald E. and D'Amico, Giovanni and Coops, Nicholas C. and Hermosilla, Txomin and White, Joanne C. and Wulder, Michael A. and Marchetti, Marco and Mugnozza, Giuseppe Scarascia and Chirici, Gherardo},
  year = 2022,
  month = feb,
  journal = {International Journal of Applied Earth Observation and Geoinformation},
  volume = {106},
  pages = {102663}
}

@article{Fuentes2025,
  title = {Vegetation browning as an indicator of drought impact and ecosystem resilience},
  author = {Fuentes, Ignacio and Lopatin, Javier and Galleguillos, Mauricio and McPhee, James},
  year = 2025,
  journal = {Science of Remote Sensing},
  volume = {11},
  pages = {100219}
}

@misc{Ginzler2021,
  title = {Vegetation Height Model {NFI}},
  author = {Ginzler, Christian},
  year = 2021,
  howpublished = {National Forest Inventory, \url{https://www.envidat.ch/dataset/vegetation-height-model-nfi}} 
}

@article{Gong2024,
  title = {Satellite remote sensing of vegetation phenology: Progress, challenges, and opportunities},
  shorttitle = {Satellite remote sensing of vegetation phenology},
  author = {Gong, Zheng and Ge, Wenyan and Guo, Jiaqi and Liu, Jincheng},
  year = 2024,
  month = nov,
  journal = {ISPRS Journal of Photogrammetry and Remote Sensing},
  volume = {217},
  pages = {149--164}
}

@inproceedings{He2016b,
  title = {Deep Residual Learning for Image Recognition},
  booktitle = {IEEE Conference on Computer Vision and Pattern Recognition},
  author = {He, Kaiming and Zhang, Xiangyu and Ren, Shaoqing and Sun, Jian},
  year = 2016,
  month = jun,
  pages = {770--778},
  publisher = {IEEE}
}

@article{Hermann2023,
  title = {Meteorological history of low-forest-greenness events in {Europe} in 2002--2022},
  author = {Hermann, Mauro and R{\"o}thlisberger, Matthias and Gessler, Arthur and Rigling, Andreas and Senf, Cornelius and Wohlgemuth, Thomas and Wernli, Heini},
  year = 2023,
  month = mar,
  journal = {Biogeosciences},
  volume = {20},
  number = {6},
  pages = {1155--1180}
}

@incollection{Hoegh-Guldberg2018,
  title = {Impacts of 1.5$^\circ${C} Global Warming on Natural and Human Systems},
  booktitle = {Global Warming of 1.5$^\circ${C}. {IPCC} Special Report},
  author = {{Hoegh-Guldberg}, Ove and Jacob, Daniela and Taylor, Michael and Bindi, Marco and Brown, Sally and Camilloni, Ines and Diedhiou, Arona and Djalante, Riyanti and Ebi, Kristie L and Engelbrecht, Francois and Guiot, Jo{\"e}l and Hijioka, Yasuaki and Mehrotra, Shagun and Payne, Antony and Seneviratne, Sonia I and Thomas, Adelle and Warren, Rachel and Zhou, Guangsheng and others},
  year = 2018,
  pages = {175--311},
  publisher={IPCC Secretariat}
}

@article{Jiang2023a,
  title = {Accuracy and consistency of space-based vegetation height maps for forest dynamics in alpine terrain},
  author = {Jiang, Yuchang and R{\"u}etschi, Marius and Garnot, Vivien Sainte Fare and Marty, Mauro and Schindler, Konrad and Ginzler, Christian and Wegner, Jan D.},
  year = 2023,
  month = dec,
  journal = {Science of Remote Sensing},
  volume = {8},
  pages = {100099},
  issn = {2666-0172},
  doi = {10.1016/j.srs.2023.100099},
  urldate = {2025-11-16}
}

@article{Klosterman2014,
  title = {Evaluating remote sensing of deciduous forest phenology at multiple spatial scales using PhenoCam imagery},
  author = {Klosterman, S. T. and Hufkens, K. and Gray, J. M. and Melaas, E. and Sonnentag, O. and Lavine, I. and Mitchell, L. and Norman, R. and Friedl, M. A. and Richardson, A. D.},
  year = 2014,
  journal = {Biogeosciences},
  volume = {11},
  number = {16},
  pages = {4305--4320}
}

@misc{Koch2024a,
  title = {Tree species map of {Switzerland}},
  author = {Koch, Tiziana and Hobi, Martina and Morsdorf, Felix and Waser, Lars},
  year = 2024,
  howpublished = {\url{https://www.envidat.ch/dataset/tree-species-map-of-switzerland}}
}

@article{Koch2025,
  title = {Assessment of tree species specific phenology metrics from {Sentinel-2} data to complement in situ monitoring},
  author = {Koch, Tiziana L. and Grubinger, Samuel and Coops, Nicholas C. and Damm, Alexander and Morsdorf, Felix and Waser, Lars T. and Wegner, Jan D. and Hobi, Martina L.},
  year = 2025,
  journal = {Ecological Indicators},
  volume = {180},
  pages = {114299}
}

@article{Koenker1978,
  title = {Regression Quantiles},
  author = {Koenker, Roger and Bassett, Gilbert},
  year = 1978,
  journal = {Econometrica},
  volume = {46},
  number = {1},
  eprint = {1913643},
  pages = {33--50}
}

@article{Kosczor2022,
  title = {Assessing land surface phenology in Araucaria-Nothofagus forests in Chile with Landsat 8/Sentinel-2 time series},
  author = {Kosczor, E. and Forkel, M. and Hern{\'a}ndez, J. and Kinalczyk, D. and Pirotti, F. and Kutchartt, E.},
  year = 2022,
  month = aug,
  journal = {International Journal of Applied Earth Observation and Geoinformation},
  volume = {112},
  pages = {102862},
  issn = {15698432},
  doi = {10.1016/j.jag.2022.102862},
  urldate = {2026-03-19},
  langid = {english}
}

@article{Liu2023a,
  title = {Vegetation browning: global drivers, impacts, and feedbacks},
  author = {Liu, Qiuyu and Peng, Changhui and Schneider, Robert and Cyr, Dominic and Liu, Zelin and Zhou, Xiaolu and Du, Mingxi and Li, Peng and Jiang, Zihan and McDowell, Nate G. and Kneeshaw, Daniel},
  year = 2023,
  journal = {Trends in Plant Science},
  volume = {28},
  number = {9},
  pages = {1014--1032}
}

@article{Loshchilov2019,
  title = {Decoupled Weight Decay Regularization},
  author = {Loshchilov, Ilya and Hutter, Frank},
  year = 2019,
  journal = {preprint arXiv:1711.05101}
}

@article{Low2020,
  title = {Phenology Modelling and Forest Disturbance Mapping with {Sentinel-2} Time Series in {Austria}},
  author = {L{\"o}w, Markus and Koukal, Tatjana},
  year = 2020,
  month = jan,
  journal = {Remote Sensing},
  volume = {12},
  number = {24},
  pages = {4191}
}

@article{Maaten2008,
  title = {Visualizing Data using {t-SNE}},
  author = {van der Maaten, Laurens and Hinton, Geoffrey},
  year = 2008,
  journal = {Journal of Machine Learning Research},
  volume = {9},
  number = {86},
  pages = {2579--2605}
}

@article{Molnar2024,
  title = {Forest Disturbance Monitoring Using Cloud-Based {Sentinel-2} Satellite Imagery and Machine Learning},
  author = {Moln{\'a}r, Tam{\'a}s and Kir{\'a}ly, G{\'e}za},
  year = 2024,
  journal = {Journal of Imaging},
  volume = {10},
  number = {1},
  pages = {14}
}

@article{OCallaghan1984,
  title = {The extraction of drainage networks from digital elevation data},
  author = {O'Callaghan, John F. and Mark, David M.},
  year = 1984,
  month = dec,
  journal = {Computer Vision, Graphics, and Image Processing},
  volume = {28},
  number = {3},
  pages = {323--344},
  issn = {0734-189X},
  doi = {10.1016/S0734-189X(84)80011-0},
  urldate = {2025-11-17}
}

@inproceedings{Pasquarella2023,
  title = {Comprehensive Quality Assessment of Optical Satellite Imagery Using Weakly Supervised Video Learning},
  booktitle = {IEEE/CVF Conference on Computer Vision and Pattern Recognition},
  author = {Pasquarella, Valerie J. and Brown, Christopher F. and Czerwinski, Wanda and Rucklidge, William J.},
  year = 2023,
  pages = {2125--2135}
}

@misc{Price2024,
  title = {The Habitat Map of {Switzerland} v1\_1 2024},
  author = {Price, Bronwyn and Kolecka, Natalia and Huber, Nica and R{\"u}etschi, Marius and Nussbaumer, Anita and Ginzler, Christian},
  year = 2024,
  howpublished = {\url{https://envidat.ch/#/metadata/the-habitat-map-of-switzerland-v1-1}},
}

@article{Reinosch2025,
  title = {Detailed validation of large-scale {Sentinel}-2-based forest disturbance maps across {Germany}},
  author = {Reinosch, Eike and Backa, Julian and Adler, Petra and Deutscher, Janik and Eisnecker, Philipp and Hoffmann, Karina and Langner, Niklas and Puhm, Martin and R{\"u}etschi, Marius and Straub, Christoph and Waser, Lars T and Wiesehahn, Jens and Oehmichen, Katja},
  year = 2025,
  journal = {Forestry: An International Journal of Forest Research},
  volume = {98},
  number = {3},
  pages = {437--453}
}

@article{Richardson2013,
  title = {Climate change, phenology, and phenological control of vegetation feedbacks to the climate system},
  author = {Richardson, Andrew D. and Keenan, Trevor F. and Migliavacca, Mirco and Ryu, Youngryel and Sonnentag, Oliver and Toomey, Michael},
  year = 2013,
  journal = {Agricultural and Forest Meteorology},
  volume = {169},
  pages = {156--173}
}

@article{Richardson2018,
  title = {Intercomparison of phenological transition dates derived from the {PhenoCam} Dataset V1.0 and {MODIS} satellite remote sensing},
  author = {Richardson, Andrew D. and Hufkens, Koen and Milliman, Tom and Frolking, Steve},
  year = 2018,
  journal = {Scientific Reports},
  volume = {8},
  number = {1},
  pages = {5679}
}

@misc{Rouault2025,
  title = {{GDAL, Geospatial Data Abstraction Library}},
  author = {Rouault, Even and Warmerdam, Frank and Schwehr, Kurt and Kiselev, Andrey and Butler, Howard and {\L}oskot, Mateusz and Szekeres, Tamas and Tourigny, Etienne and Landa, Martin and Miara, Idan and Elliston, Ben and Chaitanya, Kumar and Plesea, Lucian and Morissette, Daniel and Jolma, Ari and Dawson, Nyall and Baston, Daniel and {de Stigter}, Craig and Miura, Hiroshi},
  year = 2025,
  copyright = {MIT License},
  howpublished = {Zenodo, \url{https://zenodo.org/records/17555416}}
}

@article{Sturm2022,
  title = {Satellite data reveal differential responses of {Swiss} forests to unprecedented 2018 drought},
  author = {Sturm, Joan and Santos, Maria J. and Schmid, Bernhard and Damm, Alexander},
  year = 2022,
  journal = {Global Change Biology},
  volume = {28},
  number = {9},
  pages = {2956--2978}
}

@article{Tran2023,
  title = {HP-LSP: A reference of land surface phenology from fused Harmonized Landsat and Sentinel-2 with PhenoCam data},
  shorttitle = {HP-LSP},
  author = {Tran, Khuong H. and Zhang, Xiaoyang and Ye, Yongchang and Shen, Yu and Gao, Shuai and Liu, Yuxia and Richardson, Andrew},
  year = 2023,
  month = oct,
  journal = {Scientific Data},
  volume = {10},
  number = {1},
  pages = {691},
  publisher = {Nature Publishing Group},
  issn = {2052-4463},
  doi = {10.1038/s41597-023-02605-1},
  urldate = {2025-11-14},
  copyright = {2023 The Author(s)},
  langid = {english}
}

@article{Turner2023,
  title = {Extent, Severity, and Temporal Patterns of Damage to {Cuba}'s Ecosystems following Hurricane {Irma}: {MODIS} and {Sentinel-2} Hurricane Disturbance Vegetation Anomaly ({HDVA})},
  author = {Turner, Hannah C. and Galford, Gillian L. and Hernandez Lopez, Norgis and Falc{\'o}n M{\'e}ndez, Armando and {Borroto-Escuela}, Daily Yanetsy and Hern{\'a}ndez Ramos, Idania and {Gonz{\'a}lez-D{\'i}az}, Patricia},
  year = 2023,
  month = jan,
  journal = {Remote Sensing},
  volume = {15},
  number = {10},
  pages = {2495}
}

@article{Vitasse2009,
  title = {Leaf phenology sensitivity to temperature in {European} trees: Do within-species populations exhibit similar responses?},
  shorttitle = {Leaf phenology sensitivity to temperature in European trees},
  author = {Vitasse, Yann and Delzon, Sylvain and Dufr{\^e}ne, Eric and Pontailler, Jean-Yves and Louvet, Jean-Marc and Kremer, Antoine and Michalet, Richard},
  year = 2009,
  month = may,
  journal = {Agricultural and Forest Meteorology},
  volume = {149},
  number = {5},
  pages = {735--744}
}

@article{Vitasse2011,
  title = {Assessing the effects of climate change on the phenology of {European} temperate trees},
  author = {Vitasse, Yann and Fran{\c c}ois, Christophe and Delpierre, Nicolas and Dufr{\^e}ne, Eric and Kremer, Antoine and Chuine, Isabelle and Delzon, Sylvain},
  year = 2011,
  month = jul,
  journal = {Agricultural and Forest Meteorology},
  volume = {151},
  number = {7},
  pages = {969--980}
}

@article{Vitasse2021,
  title = {Impact of microclimatic conditions and resource availability on spring and autumn phenology of temperate tree seedlings},
  author = {Vitasse, Yann and Baumgarten, Frederik and Zohner, Constantin M. and Kaewthongrach, Rungnapa and Fu, Yongshuo H. and Walde, Manuel G. and Moser, Barbara},
  year = 2021,
  journal = {New Phytologist},
  volume = {232},
  number = {2},
  pages = {537--550}
}

@misc{Waser2025,
  title = {Forest Type {NFI}},
  author = {Waser, Lars and Ginzler, Christian and Psomas, Achilleas and R{\"u}etschi, Marius and Rehush, Nataliia},
  year = 2025,
  howpublished = {National Forest Inventory, \url{https://www.envidat.ch/dataset/forest-type-nfi}}
}

@article{Zhang2003,
  title = {Monitoring vegetation phenology using {MODIS}},
  author = {Zhang, Xiaoyang and Friedl, Mark A. and Schaaf, Crystal B. and Strahler, Alan H. and Hodges, John C. F. and Gao, Feng and Reed, Bradley C. and Huete, Alfredo},
  year = 2003,
  month = mar,
  journal = {Remote Sensing of Environment},
  volume = {84},
  number = {3},
  pages = {471--475}
}
	\end{spacing}
}


\end{document}